\documentclass{iaus}
\usepackage{graphicx}

\title[Progress in NLTE calculations] 
{Review: progress in NLTE calculations and their application to large data-sets}

\author[L. Mashonkina]   
{Lyudmila Mashonkina$^1$
}

\affiliation{$^1$ Institute of Astronomy, Russian Academy of Sciences, \\Pyatnitskaya st. 48, RU-119017 Moscow, Russia\\ email: {\tt lima@inasan.ru}  
}

\pubyear{2014}
\volume{298}  
\pagerange{XX -- YY}
\jname{IAUS\,298 Setting the scence for Gaia and LAMOST}
\editors{S. Feltzing, G. Zhao, N.\,A. Walton \& P.\,A. Whitelock, eds.}
\begin{document}

\newcommand{\avr}{$\ensuremath{\langle\mathrm{3D}\rangle}$}
\newcommand{\kH}{$S_{\!\rm H}$}    
\newcommand{\Teff}{T_{\rm eff}}
\newcommand{\Eexc}{E_{\rm exc}}
\newcommand{\eps}[1]{\log\varepsilon_{\rm #1}}

\maketitle

\begin{abstract}
One of the major tasks in interpretation of data from large-scale stellar surveys is to determine the fundamental atmospheric parameters such as effective temperature, surface gravity, and metallicity. In most on-going and upcoming projects, they are derived spectroscopically, with relying on classical one-dimensional (1D) model atmospheres and the assumption of LTE. This review discusses the present achievements and problems of non-local thermodynamic equilibrium (NLTE) line-formation calculations for FGK-type stars. The topics that are addressed include (i) the construction of comprehensive model atoms for the chemical elements with complex term system, (ii) possible systematic errors inherent in classical LTE spectroscopic determinations of stellar parameters and chemical abundances, (iii) the uncertainties in final NLTE results caused by the uncertainties in atomic data, and (iv) applications of the NLTE line-formation calculations coupled to the spatial and temporal average \avr\ models to spectroscopic analyses. 

\keywords{atomic data, line: formation, radiative transfer, stars: abundances, stars: atmospheres, stars: fundamental parameters}
\end{abstract}

\firstsection 
\section{Introduction}

Stellar elemental abundances combined with kinematics and age information serve efficiently to characterize stellar populations in our Galaxy (the Milky Way) and other galaxies. Studies of stars in the solar neighbourhood have found that the differences in elemental abundance trends between major components of the Galaxy, i.e., the thin disk, thick disk, and halo are as large as 0.3-0.5~dex and there exist substructures with differences of  0.1-0.2~dex. With the advent of large-scale stellar surveys such as RAVE, Gaia, APOGEE, LAMOST, elemental abundances will be determined for hundreds of thousands to millions of stars, spread over much larger distances than ever before. To identify different galactic populations and yet unknown components that may exist elsewhere, the stellar abundance errors should not exceed 0.1-0.2~dex. This review discusses to what extent the advances in instrumentation and observations are matched by advances in the theory.
We focus on late-type stars that dominate stellar populations in the Milky Way and dwarf galaxies surrounding it and that are suitable for investigation of the galactic chemical evolution. 

The simplest and widely used method to derive elemental abundances is based on the assumption of local thermodynamic equilibrium (LTE), where the atomic level populations at any given point in the atmosphere are calculated from the Saha and Boltzmann equations with local kinetic temperature $T$ and electron-number density. In many cases, basic stellar parameters, in particular, the surface gravity, log~g, are determined spectroscopically, also through LTE analysis. The LTE assumption is not fulfilled in spectral line formation layers, where the mean free-path of photons is not small and  the radiation field is far from TE. This review is concerned with different and more realistic NLTE approach, where the atomic level populations are determined from the balance between radiative and collisional population and de-population processes such as photoexcitation, photoionization and their inverse processes, inelastic collisions with  electrons, atoms, molecules, dielectronic recombination, charge exchange processes. For various kind particles, the velocity distribution is assumed to be Maxwellian with a single kinetic temperature $T = T_e = T_a = T_{ion}$. In NLTE, the excitation and ionization states of the matter are calculated from the solution of combined statistical equilibrium (SE) and radiation transfer equations and, at any depth point, depend on the physical conditions throughout the atmosphere. 

The paper is organized as follows. Progress in NLTE treatments for Fe~I-Fe~II and Ca~I-Ca~II and systematic biases in spectroscopic stellar parameters caused by neglecting the departures from LTE are discussed in Sect.~\ref{sect:stellar_parameters}. Section~\ref{sect:abundances} reviews recent NLTE abundance determinations for large stellar samples. Uncertainties in NLTE results caused by poorly known atomic data are discussed in Sect.~\ref{sect:uncertainties}. Most NLTE studies of cool stars deal with classical plane-parallel (1D) model atmospheres, because, at present, there are no tools to perform 3D-NLTE line formation calculations for atoms with a complicated term structure. The full 3D line-formation calculations differ from their 1D counterparts due to different mean atmospheric structures and the existence of atmospheric inhomogeneities. The effect of different mean atmospheric structures on emergent fluxes and derived chemical abundances can be evaluated using the spatially and temporally averaged atmospheric stratification denoted \avr. Section~\ref{sect:3d} reviews recent applications of the NLTE line-formation calculations coupled to the average \avr\ models to spectroscopic analyses. The conclusions and recommendations are given in Sect.~\ref{Conclusion}.

\section{Spectroscopic stellar parameters: NLTE vs. LTE}\label{sect:stellar_parameters}

\underline{\it Progress in NLTE calculations for iron lines.}
Iron plays an outstanding role in studies of cool stars thanks to the many lines in the visible spectrum, which are easy to detect even in very metal-poor (VMP, [Fe/H] $< -2$) stars. Hereafter, [X/H] = $\log(N_{\rm X}/N_{\rm H})_{star} - \log(N_{\rm X}/N_{\rm H})_{Sun}$. Iron lines are used to derive basic stellar parameters, and iron serves as an indicator of overall metal abundance. In stellar atmospheres with the effective temperature, $\Teff$, higher than 4500~K, neutral iron is a minority species, and its SE can easily deviate from thermodynamic equilibrium due to deviations in the ionizing radiation from the black body radiation. Therefore, since the beginning of the 1970s, a large number of studies attacked the NLTE problem for iron in the atmospheres of the Sun and cool stars (for references, see \cite[Mashonkina \etal\ 2011]{Mashonkina_fe}). It was understood that the main NLTE mechanism for Fe~I is ultra-violet (UV) overionization of the low-excitation levels. This results in weakened lines of Fe~I compared with their LTE strengths and positive NLTE abundance corrections. However, a consensus was not achieved with respect to a magnitude of the NLTE effects. 

\cite{Mashonkina_fe} suggested to include in the model atom not only the known measured energy levels of Fe~I as was done in all the previous NLTE research of iron, but also the predicted levels from atomic structure calculations of \cite{Kurucz2009}. For SE of the minority species, both low- and high-excitation levels are important. The low-excitation levels with the thresholds in the UV are subject to overionization, while the role of high-excitation levels is to compensate, in part, population loss via collisional coupling to the large continuum reservoir of the next ionization stage with a subsequent sequence of spontaneous transitions down to the lower levels. With a fairly complete model atom for Fe~I, progress was made in establishing close collisional coupling of the Fe~I levels near the continuum to the ground state of Fe~II. 
\cite{Bergemann_fe} produced very similar model atom for Fe~I, with the difference in combining the predicted energy levels. \cite{Mashonkina_fe} treated all the high-excitation predicted levels as the 6 super-levels, while \cite{Bergemann_fe} made 62 super-levels. For SE of Fe~I, rather the combined effect of the predicted energy levels than the effect of the individual levels is important. Therefore, the NLTE results obtained with model atoms of \cite{Mashonkina_fe} and \cite{Bergemann_fe} appear to be very similar, when dealing with common spectral lines, stellar parameters, and model atmospheres.

\begin{figure}[b]
\begin{center}
 \includegraphics[width=3.0in]{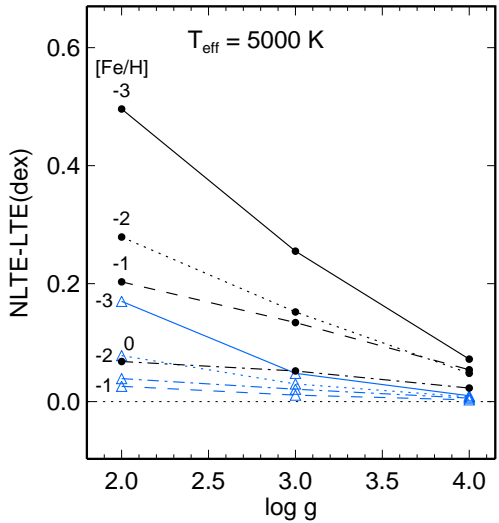} 
 \caption{NLTE abundance corrections (dex) for Fe~I 5236\,\AA\ (models with [Fe/H] = 0 and $-1$) and Fe~I 5586\,\AA\ ([Fe/H] = $-2$ and $-3$) depending on gravity and metallicity from the calculations with \kH\ = 0.1 (filled circles connected by various type lines) and \kH\ = 1 (triangles). Continuous, dotted, dashed, and dash-dotted lines correspond to the models with [Fe/H] = $-3$, $-2$, $-1$, and 0, respectively. Everywhere, $\Teff$ = 5000~K.}
   \label{ml:fig_corr} 
\end{center}
\end{figure}

Figure~\ref{ml:fig_corr} shows the NLTE abundance corrections, $\Delta_{\rm NLTE} = \eps{NLTE}-\eps{LTE}$, as computed using model atom of \cite{Mashonkina_fe}, with different treatment of inelastic collisions with H~I atoms. The hydrogen collision rates were calculated with the classical \cite[Drawin (1968, 1969)]{D68,D69} formula, and they were scaled by a factor of either \kH\ = 0.1 or \kH\ = 1. 
As expected, the departures from LTE for Fe~I grow toward lower [Fe/H] and log~g.
The NLTE corrections for lines of Fe~I were computed by \cite{Lind_fe} for an extensive grid of the MARCS model atmospheres (\cite{MARCS}) with $\Teff$ = 4000-8000~K, log~g = 1-5, and [Fe/H] = 0.5 down to $-5$. Everywhere, they adopted \kH\ = 1. It is worth noting that for the mildly metal-poor ([Fe/H] $\ge -2$) dwarfs and subgiants (log~g $\ge$ 3.5) with $\Teff \le$ 6000~K, $\Delta_{\rm NLTE}$ does not exceed 0.1~dex, independent of either \kH\ = 1 or 0.1. In each model atmosphere, the NLTE abundance corrections for different lines of Fe~I are only weakly sensitive to the excitation energy of the lower level of the transition, $\Eexc$, and the line strength.

\underline{\it Stellar gravity from iron lines.} Establishing the ionization equilibrium between Fe~I and Fe~II is a standard spectroscopic method of deriving the stellar surface gravity. \cite{Ruchtietal2013} evaluated the differences in log~g between NLTE and LTE determinations for the stellar sample selected from the RAVE survey (see their Fig.~8). Since the departures from LTE lead to weakened lines of Fe~I, but they do not affect lines of Fe~II until the extremely low metallicities, the ionization equilibrium Fe~I/Fe~II is achieved for higher gravities in NLTE than in LTE. When using common $\Teff$ in the NLTE and LTE analysis, systematic biases in gravity can be up to 0.2~dex in the metallicity range $ -2 <$ [Fe/H] $< -0.5$ and up to 0.3~dex for the most MP and most luminous stars.

\begin{figure}[b]
\begin{center}
 \includegraphics[width=2.6in]{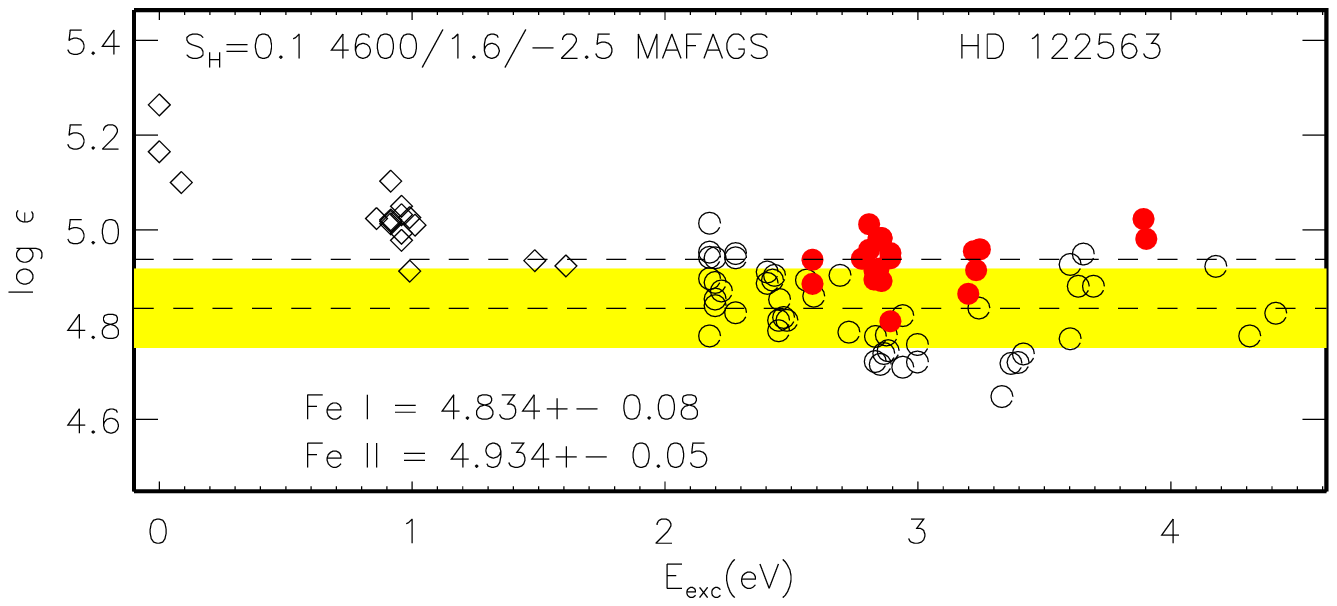} 
 \includegraphics[width=2.6in]{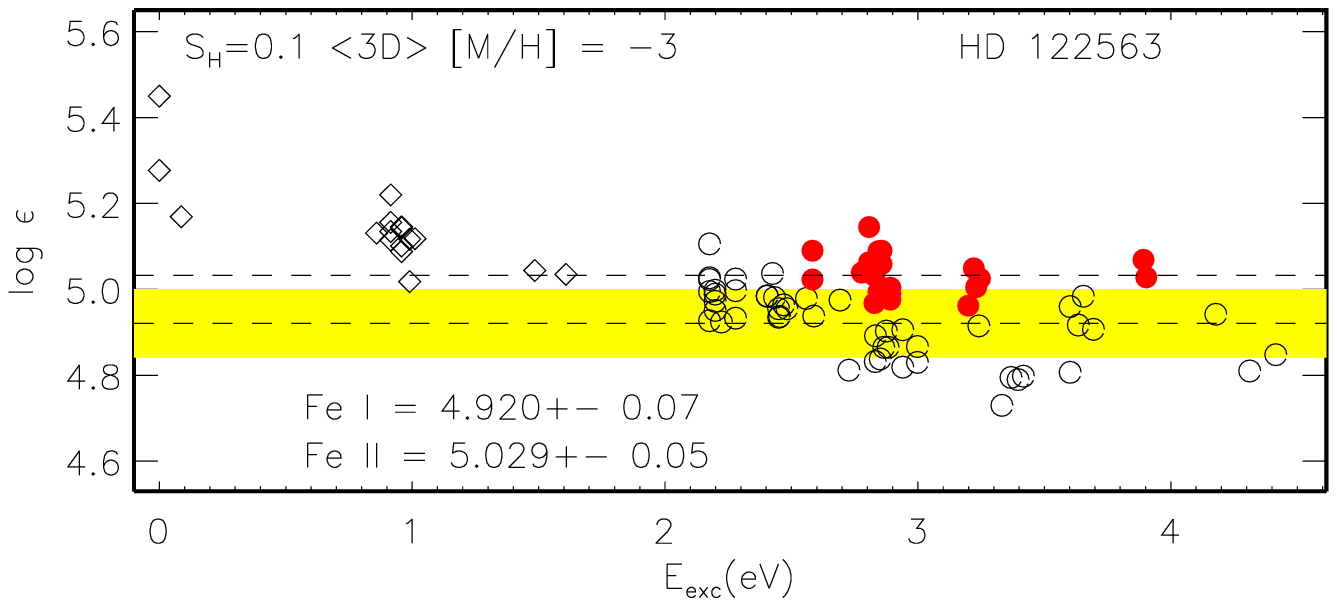} 
 \caption{NLTE (\kH\ = 0.1) abundances derived from the Fe~I (open circles for $\Eexc >$ 2~eV and rombs for $\Eexc <$ 2~eV) and Fe~II (filled circles) lines in HD\,122563 as a function of excitation energy of the lower level from the calculations with the MAFAGS-OS 4600/1.60/$-2.5$ (left panel) and ${\rm CO^5BOLD}$ \avr\ 4590/1.60/$-3$ (right panel) model atmospheres. The bottom and top dashed lines indicate the mean abundances from the Fe~I ($\Eexc >$ 2~eV) and Fe~II lines, respectively.
}
   \label{ml:fig_1D_3D} 
\end{center}
\end{figure}

\underline{\it Stellar effective temperature from iron lines.} The effective temperature can be inferred from the excitation balance of Fe~I lines, flattening abundance trends with the excitation energy of the lower level of the transition. \cite{Bergemann_fe} checked the NLTE abundances from individual lines of Fe~I as a function of $\Eexc$ in the three nearby halo stars with well determined temperatures. It was shown that the low-excitation ($\Eexc <$ 2~eV) lines reveal 0.15-0.2~dex higher abundances compared with the mean from the remaining lines of Fe~I. Very similar results were obtained by \cite{Klevasetal2013} for HD\,122563 (Fig.~\ref{ml:fig_1D_3D}), with $\Teff$ = 4600~K based on the star's angular diameter and total flux measurements. 

 One possible explanation is pronounced 3D effects for Fe~I. \cite{Collet2007} and \cite{Hayek2011} predicted that the 3D-1D abundance corrections are negative for the low-excitation lines of Fe~I and depend strongly on $\Eexc$. For example, for the line arising from the ground state in the [Fe/H] = $-1$ model atmosphere of cool giant, the 3D-1D abundance correction is 0.2~dex more negative compared with that for the $\Eexc$ = 2~eV line. The difference increases toward lower metallicity and reaches 0.6~dex at [Fe/H] = $-2$.
Thus, one should be cautious, when deriving $\Teff$ from excitation balance of the Fe~I lines. 

Substantial errors in stellar parameters can arise when both $\Teff$ and log~g 
are derived from Fe lines, through LTE analysis of excitation and ionization equilibrium, up to 400~K in $\Teff$ and 1~dex in log~g according to \cite{Ruchtietal2013}.

\underline{\it Effective temperature from the Balmer lines.}
For late type stars, the hydrogen Balmer lines can be used as indicator of $\Teff$. Applying the \cite{AG66} resonance-broadening theory, \cite[Fuhrmann \etal\ (1993, 1997)]{Fuhr93,Fuhr97} proved that the wings of H$_\alpha$ in the reference stars with well determined stellar parameters such as the Sun and Procyon are well reproduced with given $\Teff$. Later on, \cite{BPO} and \cite{Allard2008} revised the cross-sections of the self-resonance broadening of the Balmer lines by collisions with neutral hydrogen atoms and significantly improved their accuracy compared with the \cite{AG66} values. However, with the improved atomic data, the theory appears not to reproduce the observations. To fit the solar H$_\alpha$ flux profile using the MARCS models, one needs to reduce $\Teff$ of the model by 90~K compared with the solar temperature $\Teff$ = 5780~K (see Fig.~13 from \cite[Allard \etal\ 2008]{Allard2008}). Applying the \cite{Allard2008} theory and ATLAS9 model atmospheres (\cite[Kurucz 2009]{Kurucz2009}), \cite{Cayreletal2011} determined the LTE effective temperature from H$_\alpha$ for the eleven stars with an accurate $\Teff$ derived from their apparent angular diameter and found that the two sets of temperatures have a significant offset of 100~K. In NLTE, the offset could be 30-50~K smaller (\cite{Mashonkina_subaru}), but could not be removed completely.

In cool stars, wings of H$_\alpha$ are sensitive to treatment of convection in 1D model atmospheres. One can expect that a 3D approach, with radiative hydrodynamical models, could provide full understanding of the difference between the Balmer line and direct effective temperature. 
Full 3D-NLTE line formation calculations for solar H$_\alpha$ were performed for the first time by \cite{Pereiraetal2013}. As seen in their Fig.~9, the theoretical 3D profiles are still too strong compared with the solar center-disk intensity profile of H$_\alpha$. The difference between the true solar and spectroscopic effective temperature amounts to 50~K and 37~K for the 3D hydrodynamical and magneto-hydrodynamical model, respectively.

\begin{figure}[b]
\begin{center}
 \includegraphics[width=2.6in]{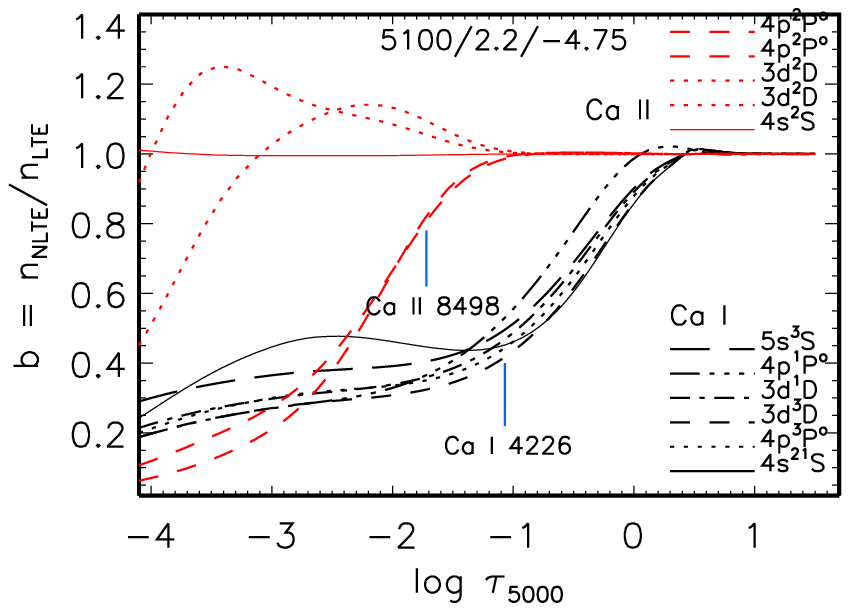} 
 \includegraphics[width=2.6in]{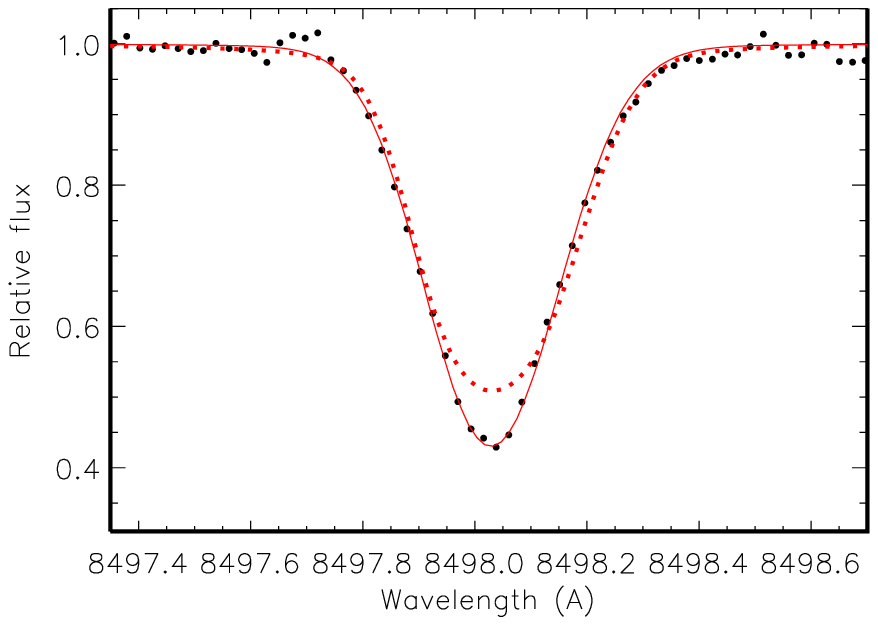} 
 \caption{Left panel: departure coefficients, $b$, for selected levels of Ca~I and Ca~II as a function of $\log \tau_{5000}$ in the model atmosphere 5100/2.2/$-4.75$. Tick marks indicate the locations of line center optical depth unity for Ca~I 4226\,\AA\ and Ca~II 8498\,\AA. Right panel: best NLTE (continuous curve) and LTE (dotted curve) fits of the Ca~II 8498\,\AA\ line in HE\,0557-4840 (bold dots). In NLTE, the Ca abundance is $\eps{Ca}$ = 2.34, and it is 0.55~dex higher in LTE.
}
   \label{ml:fig_he0557} 
\end{center}
\end{figure}

\underline{\it Gravity and metallicity of EMP stars from the Ca lines.}
One important goal of the on-going and coming stellar spectroscopic surveys is the search for the extremely metal-poor (EMP, [Fe/H] $< −3$) stars in the Milky Way and the dwarf spheroidal galaxies (dSphs) surrounding it. 
When the iron abundance falls below [Fe/H] = $-4.5$, the only chemical element visible in two ionization stages is calcium. Lines of Ca~I and Ca~II can be potent tools in deriving accurate values for surface gravity and for the Ca abundance itself.

We present recent NLTE results for HE\,0557-4840 that is the 3rd most iron-deficient star currently known, with [Fe/H] = $-4.75$. Its stellar parameters were first determined by \cite{Norrisetal2007}: $\Teff$ = 5100~K from photometric color calibrations and log~g = 2.2 from several methods, namely, an appropriate isochrone, the constraints of Balmer-profile fitting, and the NLTE analysis of the ionization equilibrium between Fe~I and Fe~II and between Ca~I and Ca~II. An accuracy of the gravity needs to be improved because the two measured lines of Fe~II are very weak and, of the two Ca~II lines, the resonance line is too strong and Ca~II 3706\,\AA\ is blended by the high members of the Balmer series. \cite{Renetal2013} took an advantage of new spectral observations in the infrared (IR) spectral range, where the Ca~II 8498, 8542, 8662\,\AA\ (IR triplet) lines are located. One of them is shown in Fig.\,\ref{ml:fig_he0557} (right panel). Note, the observed profile cannot be reproduced in LTE. The shift in wavelength between the NLTE and LTE profile is due to different NLTE effects on different isotopic components of the line. 
The departures from LTE for different lines of Ca are easy to understood from inspecting the departure coefficients, $b = n^{\rm NLTE}/n^{\rm LTE}$, for levels of Ca~I and Ca~II in the model representing atmosphere of the investigated star (Fig.\,\ref{ml:fig_he0557}, left panel). Here, $n^{\rm NLTE}$ and $n^{\rm LTE}$ are the statistical
equilibrium and thermal (Saha-Boltzmann) number densities, respectively.
For Ca~I, the main NLTE effect is the UV overionization resulting in depleted populations for all levels in the line-formation layers and, thus, weakened resonance line. The Ca~II IR triplet lines originate from the transition 3d-4p. In the layers where these lines form, photon losses in the lines themselves lead to depopulation of the upper level 4p and overpopulation of the lower level 3d. As a result, the Ca~II IR triplet lines are strongly strengthened compared with their LTE strengths. In contrast, Ca~II 3706\,\AA\ is weakened because the level 4p is the lower level of the corresponding transition. 

The NLTE abundances determined from Ca~I 4226\,\AA\ and the four lines of Ca~II appear to be consistent within the error bars, when using log~g = 2.2, i.e., $\eps{Ca I}$ = 2.22 and $\eps{Ca II}$ = 2.28$\pm$0.05. With log~g = 2.7, the difference between Ca~I and Ca~II exceeds 2$\sigma$, i.e., $\eps{Ca I}$ = 2.26 and $\eps{Ca II}$ = 2.43$\pm$0.07. These results support the earlier determination of gravity of HE\,0557-4840 and significantly improve its accuracy.

\underline{\it Ca~II triplet lines as metallicity indicator.}
The Ca II triplet lines (CaT) are used for metallicity estimates of distant individual red giant stars and integrated stellar populations. Their advantage is that they remain very strong and broad until very low element abundances and they can be measured with sufficient accuracy at a moderate resolution. For decades, one employed an empirical relation between [Fe/H] and equivalent width of CaT that has been calibrated on large samples of individual reg giant stars in globular clusters. However, having applied this technique to individual stars in the classical dSphs, one found the significant lack of EMP stars. Based on NLTE calculations of \cite[Mashonkina \etal\ (2007)]{Mashonkina_ca} for Ca~I-Ca~II, \cite{Starkenburgetal2010} proposed to revise the CaT-[Fe/H] relation in the low-metallicity regime by taking substantial departures from LTE for the Ca~II triplet lines into account. The new CaT calibration is valid for $-0.5 \ge$ [Fe/H] $\ge -4$, as proved by the comparison with the [Fe/H] values as determined from high-resolution measurements for the stars in the galaxies Sculptor, Fornax, Carina, and Sextans. It appears, the classical dSphs are not devoid of EMP stars. 

\cite{Spiteetal2012} confirmed that the NLTE abundances can be reliably derived from the Ca~II IR triplet lines in the wide range of metallicities. They found a mean abundance difference of 0.07~dex between subordinate lines of Ca~I and Ca~II 8662\,\AA\ for the 53 stars from the Large Programme “First Stars”. 

\section{Stellar elemental abundances: NLTE vs. LTE}\label{sect:abundances}

NLTE abundances can be determined for many chemical elements in late-type stars, namely, Li-O, Na-Si, S, K-Ti, Cr-Ni, Zn, Sr, Zr, Ba, Pr, Eu, Pb, and Th. For the NLTE methods treated before 2005, a thorough review of \cite{Asplund05} can be recommended. In recent years, the first NLTE studies were performed for Sc~I-II (\cite{Zhang_sc}), Ti~I-II (\cite{Bergemann_ti}), Cr~I-II (\cite[Bergemann \& Cescutti 2010]{Bergemann_cr}), Mn~I-II (\cite{Bergemann_mn}), Co~I-II (\cite{Bergemann_co}), Ni~I (\cite{nlte_ni1}), Zn~I (\cite{Takedaetal05}), Zr~I-II (\cite{Velichko_zr}), Pr~II (\cite{Mashonkina_pr}), Pb~I and Th~II (\cite{Mashonkina_pb_th}). 
This review cannot cover all the interesting NLTE results, and only few of them are described. 

\underline{\it NLTE abundances for the Large Programme “First Stars”, PI: R. Cayrel.} In 2007-2012, the NLTE abundances of Na, Mg, Al, S, K, Ca, Sr, and Ba were determined for stellar samples of \cite{Cayreletal2004} and \cite{Bonifacioetal2009}. The abundance ratios between the odd-Z and even-Z elements, such as Na/Mg and Al/Mg, are particularly important for testing the nucleosynthesis models. Exactly these elemental ratios are most affected by the departures from LTE. As shown by \cite[Andrievsky \etal\ (2007, 2008)]{Andrievsky_na1,Andrievsky_al1}, the [Na/Mg] abundance ratios obtained under the LTE assumption can be larger  by up to 0.5~dex, while the [Al/Mg] ratios can be smaller by up to 0.6~dex compared with the corresponding NLTE values, and the departures from LTE are very sensitive to stellar surface gravity. NLTE leads to consistent elemental ratios between dwarfs and giants that should be considered as favoring NLTE line formation.

\underline{\it Cr abundances of VMP stars.} 
Classical LTE abundance analyses reveal underabundance of chromium relative to iron at low metallicities, where Cr is mostly observed in lines of Cr~I. Similarly to Fe~I, Cr~I is a minority species in the atmospheres with $\Teff >$ 4500~K, and it is expected to be subject to the UV overionization. \cite{Bergemann_cr} have shown that NLTE leads to weakened lines of Cr~I and positive NLTE abundance corrections of up to +0.3~dex in the [Fe/H] $\simeq -2.5$ stars. For the Sun and small sample of MP stars, the NLTE abundances from the two ionization stages, Cr~I and Cr~II, appear to be consistent, and the Cr abundance follows the Fe one, with [Cr/Fe] $\simeq$ 0. This gives a strong argument for common origin of Cr and Fe.

\underline{\it NLTE vs. LTE for the Gaia lines of Mg~I and Ca~I-II.} 
\cite{Merleetal2011} evaluated the NLTE effects for lines of Mg~I and Ca~I-II in the grid of model atmospheres of cool giants with 3500~K $\le \Teff \le$ 5250~K, 0.5 $\le$ log~g $\le$ 2, and $-4 \le$ [Fe/H] $\le$ 0.5. They focused on the IR lines important for the Gaia project. The results were presented as equivalent width ratios between NLTE and LTE, $W_{\rm NLTE}/W_{\rm LTE}$. One should be very cautious, when applying such ratios to stellar abundance analysis. For example, in the models with $\Teff$ = 5000~K and log~g = 2, a 10\%\ variation in equivalent width of Mg~I 8736\,\AA\ leads to 0.04~dex variation in derived element abundance, when [Fe/H] = $-2$, but larger abundance variations of 0.09~dex and 0.14~dex for [Fe/H] = $-1$ and 0, respectively. This is due to moving the line from the linear to saturated part of the curve of growth with increasing metallicity (element abundance).

Another concern is ignoring inelastic collisions with H~I atoms in NLTE calculations of \cite{Merleetal2011}. Using model atom for Mg~I from \cite{Mashonkina_mg}, I checked the influence of different treatment of collisional processes on NLTE results for Mg~I 8736\,\AA\ in the 5000/2/$-2$ model. With pure electronic collisions, $\Delta_{\rm NLTE}$ = +0.17~dex, in line with \cite{Merleetal2011}. However, the NLTE abundance correction amounts to $-0.04$~dex and +0.05~dex, when including H~I collisions with accurate data from \cite{mg_hyd2012} and with the Drawinian rates scaled by \kH\ = 0.1, respectively.

\section{Uncertainties in NLTE studies}\label{sect:uncertainties}

NLTE calculations for given atom require a complete set of atomic data on energy levels, photoionization cross-sections, transition probabilities, collision excitation and collision ionization cross-sections. For elements up to Ca and Fe, the situation with accuracy of radiative data has been greatly improved due to realization of the Opacity Project and IRON project, with the data available through the TOPbase\footnote{\tt http://cdsweb.u-strasbg.fr/topbase/topbase.html} and TIPbase\footnote{\tt http://cdsweb.u-strasbg.fr/tipbase/home.html} database. 

The main source of uncertainties in NLTE results is poorly known collisional processes. 
  Among all particles the collision frequency is the highest for electrons because they have the highest velocities. However, in cool atmospheres, where the matter is primarily neutral, inelastic collisions with neutral hydrogen atoms can be numerous and important. Accurate data from quantum mechanical calculations of collisions with H~I atoms are available for Li~I (\cite{BelyaevBarklem2003}), Na~I (\cite{Belyaevetal2010}), and Mg~I (\cite[Barklem \etal\ 2012]{mg_hyd2012}, hereafter, BBSGF). Most NLTE studies should rely on the classical \cite[Drawin (1968, 1969)]{D68,D69} formula as implemented by \cite{Steenbock1984}. 
   The Drawin formalism was criticized for not providing a realistic description of the physics involved. However, in the absence of more reliable calculations, the only possibility to obtain an indication of the influence of H~I collisions is to adopt the Drawinian rates. In the SE calculations, they are scaled by a factor \kH\ which is estimated empirically for each chemical species from different influence of collisions on its different lines in different objects.

When applying the Drawin's formula to Li~I and Na~I, the collision rates appear to be strongly overestimated compared with the accurate data, by up to 4 orders of magnitude. For Mg+H collisions, \cite{mg_hyd2012} showed that the accurate rates for excitation and ion-pair production processes are larger than those for Li~I and Na~I.
Using the BBSGF data,
\cite{Mashonkina_mg} found that Mg+H collisions serve as an efficient thermalizing process for the SE of Mg~I in the atmospheres of MP stars and Mg abundance determinations for the two representative halo stars HD\,84937 (6350/4.09/$-2.08$) and HD\,122563 (4600/1.60/$-2.5$) are improved, though the differences between different lines are not completely removed.  

The changes in collisional rates between BBSGF and Drawin's recipe result in different effects for different lines in different stars. For example, for Mg~I 4571, 4703, and 5528\,\AA\ in the 6350/4.09/$-2.08$ model, \cite{Mashonkina_mg} obtained very similar results, independent of either the accurate Mg+H collision data or the Drawinian rates scaled by \kH\ = 0.1 were applied. For the VMP cool giant model, the departures from LTE are greater when applying the BBSGF rates. However, nowhere does the difference between using these two recipes exceed 0.1~dex. This is smaller than or similar to a line-to-line scatter for stellar abundance determinations. 

\section{\avr-NLTE: what is advantage for spectroscopic analyses?}\label{sect:3d}

The final point that is addressed in this review is to what extent the NLTE calculations coupled to the average \avr\ models are useful in situations where 3D effects are pronounced.

\cite{Mashonkinaetal2013} checked whether the \avr-NLTE line-formation calculations for Fe~I can reproduce observations of the centre-to-limb variation of the solar lines. 
The two Fe~I lines at 7780\,\AA\ and 6151\,\AA\ were selected from measurements of \cite{Pereira2009}. The first one is weakened toward the limb, and the second one is, in contrast, strengthened toward the limb. It was shown that (i) for both lines, the classical MAFAGS-OS solar model atmosphere (\cite{Grupp09}) predicts decreased equivalent widths toward the limb and too low values of $W_\lambda$ compared with the observed ones at the limb; (ii) the NLTE calculations with the average \avr\ model obtained from the full 3D ${\rm CO^5BOLD}$ solar model (\cite{Freytag2012}) reproduce nearly perfectly the behavior of Fe~I 7780\,\AA, but the average model is unsuccessful in the case of Fe~I 6151\,\AA, both in LTE and NLTE; (iii) full 3D-LTE calculations of Fe~I 6151\,\AA\ correctly predict the growth of $W_\lambda$ toward the limb, but overestimate theoretical values. The latter can be due to neglecting the departures from LTE. 
   These results illustrate the varying sensitivity of different Fe~I lines to 3D effects. 
The Fe~I 7780\,\AA\ line is mainly affected by the change in mean atmospheric structure of 3D compared with 1D model. In such a case, spectroscopic analysis benefits from using \avr\ models. Fe~I 6151\,\AA\ is more sensitive to atmospheric inhomogeneity that is not included in \avr\ models.

\cite{Bergemann_fe} applied the \avr\ models from calculations with the Stagger code to derive the NLTE abundances from individual lines of Fe~I and Fe~II in the three VMP stars. For one common star, HD\,122563, \cite{Klevasetal2013} performed similar analysis, but using the ${\rm CO^5BOLD}$ \avr\ model 4590/1.6/$-3$ (Fig.\,\ref{ml:fig_1D_3D}, right panel). Though the Stagger model of HD\,122563 is 75~K hotter, this cannot explain the differences in abundances obtained with the Stagger and ${\rm CO^5BOLD}$ model for this star. In the ${\rm CO^5BOLD}$ model, lines of Fe~I reveal excitation trend, with $\Delta\eps{}$(($\Eexc <$ 2eV) -- ($\Eexc >$ 2eV)) = 0.22~dex. This is similar to that found with the 1D models (see Sect.\,\ref{sect:stellar_parameters}). \cite{Bergemann_fe} obtained smaller abundance difference of 0.08~dex between the low- and high-excitation lines, however, the difference exists. The departures from LTE for Fe~I are larger in the Stagger \avr\ model, with the mean $\Delta_{\rm NLTE}$ = +0.14~dex (with \kH\ = 1), compared with those in the ${\rm CO^5BOLD}$ model, where $\Delta_{\rm NLTE}$ = +0.08~dex with \kH\ = 0.1. Despite this fact, \cite{Bergemann_fe} obtained more negative difference Fe~I ($\Eexc >$2eV) -- Fe~II = $-0.16$~dex compared with $-0.11$~dex in the \cite{Klevasetal2013} study.
The difference between 3D models of HD\,122563 computed with the Stagger and ${\rm CO^5BOLD}$ code is evident from the LTE analysis, too. For Fe~I, the LTE abundance difference (Stagger \avr\ -- MARCS) = $-0.07$~dex (\cite[Bergemann \etal\ 2012]{Bergemann_fe}), while (${\rm CO^5BOLD}$ \avr\ -- MARCS) = +0.25~dex (\cite[Klevas \etal\ 2013]{Klevasetal2013}). To be applied to stellar spectroscopic analyses, the 3D models need to be further tested, in particular, for MP atmospheres. 

The \avr-NLTE analysis of Fe~I/Fe~II in the reference VMP stars with well determined atmospheric parameters does not provide any advantage compared with the 1D-NLTE one. \cite[Bergemann \etal\ (2012, Table~3)]{Bergemann_fe} obtained the following differences in abundances between Fe~I and Fe~II using the MARCS 1D and Stagger \avr\ models.

\begin{table}[htbp]
\begin{tabular}{llrlr}
HD~~~84937, 6408/4.13/$-2.16$, & 1D-NLTE: & 0.06~dex, & \avr-NLTE: & 0.07~dex, \\
HD~140283, 5777/3.70/$-2.38$, &           & 0.06~dex, &            & $-0.04$~dex, \\
HD~122563, 4665/1.64/$-2.51$, &           & $-0.08$~dex, &         & $-0.16$~dex. \\
\end{tabular}
\end{table}

\section{Conclusions and recommendations}\label{Conclusion}


\underline{\it Spectroscopic stellar parameters.}

{\it Effective temperature.} At present, no spectroscopic method can be recommended for reliable determinations of $\Teff$ for cool stars, in particular, MP stars. 
The method of excitation equilibrium of Fe~I is not working, when employing the 1D model atmospheres, even with NLTE line formation (\cite[Mashonkina \etal\ 2011]{Mashonkina_fe}, \cite[Bergemann \etal\ 2012]{Bergemann_fe}, \cite[Klevas \etal\ 2013]{Klevasetal2013}). Effective temperature determined from the Balmer line H$_{\alpha}$ is systematically underestimated, by about 100~K in the 5100~K $\le \Teff \le$ 6500~K, log~g $\ge$ 3.9, and $-0.7 \le$ [Fe/H] $\le$ 0.2 stellar parameter range, according to 1D-LTE analysis of \cite{Cayreletal2011}. Full 3D-NLTE line-formation calculations by \cite{Pereiraetal2013} for H~I have improved the situation with determination of $\Teff$(H$_{\alpha}$) for the Sun, and further tests are needed for MP stars.

{\it Surface gravity.} The method of ionization equilibrium between Fe~I and Fe~II and between Ca~I and Ca~II can be successfully applied to derive surface gravity of cool stars provided that the departures from LTE are taken into account. The main source of uncertainties in NLTE calculations is poorly known inelastic collisions with H~I atoms. To obtain appropriate results from Ca~I/Ca~II, the \cite[Drawin (1968, 1969)]{D68,D69} collisional rates have to be reduced by a factor of 10 (\cite{Mashonkina_ca}). For Fe~I/Fe~II, the scaling factor appears different for MP dwarfs, with \kH\ = 1, and giants, with \kH\ = 0.1 (\cite[Mashonkina \etal\ 2011, 2013]{Mashonkina_fe,Mashonkinaetal2013} and \cite[Bergemann \etal\ 2012]{Bergemann_fe}). 

For the problem of stellar surface gravities, the coming Gaia parallaxes are of extreme importance. They will also help to constrain/test collisional data for Ca~I and Fe~I.

\underline{\it Stellar elemental abundances.}
Pronounced departures from LTE are found for most chemical species at low metallicities. For atoms in their minority ionization stage, the NLTE abundance corrections can be as large as 0.2~dex even for weak spectral lines. Therefore, an adequate line formation modelling is of the first priority for stellar abundance determinations in the galactic chemical evolution studies, where stellar sample covers 2-4 orders of magnitude in metallicity. 

\underline{\it \avr-NLTE line-formation modelling for Fe~I.}

Recent NLTE line-formation calculations for Fe~I coupled to the average \avr\ models show that they are of limited applicability (\cite[Mashonkina \etal\ 2013]{Mashonkinaetal2013}, \cite[Klevas \etal\ 2013]{Klevasetal2013}).

Quite well agreement of results from 3D radiative hydrodynamic simulations of convection near the {\it solar} surface generated with the ${\rm CO^5BOLD}$ and Stagger codes was obtained by \cite{3Dcodes}. As shown above, the situation seems to be different in case of the VMP cool giant HD\,122563. Further tests of 3D models are needed for MP atmospheres.


To summarize, the theory was advanced in recent years, however, much more should be done to match advances in instrumentation and observations.

{\it Acknowledgements.}
The author acknowledges a partial support from the International Astronomical Union and the Presidium RAS Programme ``Non-stationary phenomena in the Universe'' of the participation at the IAUS~298.

\begin{discussion}


\discuss{H.-G. Ludwig}{When you showed the NLTE and LTE profiles of Ca~II 8498\,\AA\ (Fig.\,\ref{ml:fig_he0557}) you mentioned that the NLTE effects are different for different isotopic components. For clarification: is this only due to line strength, or are these effects related to the atomic structure?}


\discuss{L. Mashonkina's answer}{For SE calculations, calcium can be considered as a single isotope because the difference in energy between common levels in different isotopes of Ca~I and Ca~II does not exceed 3$\cdot 10^{-5}$~eV and their populations should be in thermal equilibrium to each other. The difference in NLTE effects between different isotopic components of Ca~II 8498\,\AA\ is fully determined by different line strengths. The isotope $^{40}$Ca is approximately one and a half order of magnitude more abundant compared with $^{44}$Ca that is the second most populated isotope of Ca. Therefore, line of $^{40}$Ca is the strongest and has the largest departures from LTE among six isotopic components that results in shifting the synthetic NLTE profile in the short-wave range relative to the LTE one.}

\discuss{P. E. Nissen}{How are the prospects to the full 3D-NLTE calculations for different elements?}

\discuss{L. Mashonkina's answer}{At present, self-consistent 3D-NLTE line-formation calculations can be performed for Li~I (Barklem \etal\ 2003, A\&A, 409, L1; Cayrel \etal\ 2007, A\&A, 473, L37), O~I (Asplund \etal\ 2004, A\&A, 417, 751), and H~I (Pereira \etal\ 2013, arxiv: 1304.7679). Next suitable candidates are, probably, Na~I, Mg~I, and Ca~I, with the atomic term structure that can be represented by model atom with 15-20 levels.}

\end{discussion}

\end{document}